\documentclass[11pt]{article} 
\usepackage{jcappub,natbib,float,caption,comment,bm}

\usepackage[normalem]{ulem}

\title{Constraints on hidden photons from current and future observations of CMB spectral distortions}
\author[a]{Kerstin E. Kunze,}
\affiliation[a]{Departamento de F\'\i sica Fundamental and IUFFyM,
Universidad de Salamanca, Plaza de la Merced s/n, 37008 Salamanca,
Spain}
\author[a]{Miguel \'A. V\'azquez-Mozo}
\emailAdd{kkunze@usal.es}
\emailAdd{Miguel.Vazquez-Mozo@cern.ch}
\abstract{
A variety of beyond the standard model scenarios contain very light hidden sector U(1) gauge bosons 
undergoing kinetic mixing with the photon. The resulting oscillation between ordinary and hidden photons 
leads to spectral distortions of the cosmic microwave background. We update the bounds on 
the mixing parameter $\chi_0$ and the mass of the hidden photon $m_{\gamma'}$ for future experiments  measuring 
CMB spectral distortions, such as PIXIE and PRISM/COrE.
For  $10^{-14}\;{\rm eV}\lesssim m_{\gamma'}\lesssim 10^{-13}\;{\rm eV}$, we find
the kinetic mixing angle $\chi_0$ has to be less
than $10^{-8}$ at 95\% CL.  These bounds are  more than an order of magnitude stronger than 
those derived from the COBE/FIRAS data.
}
\begin{document}
\maketitle

\section{Introduction}
\setcounter{equation}{0}

The search for experimental signatures of beyond the standard model (BSM) physics is hindered by the small size of 
low energy effects. On general grounds, high energy dynamics is codified in higher dimensional operators
in the effective action, whose couplings are suppressed by inverse powers of the energy scale of new physics.  
As a consequence, corrections to the standard model results are very small when not completely negligible. 

A different situation might occur when integrating out a massive field gives rise to
a marginal operator in the effective action whose coupling depends logarithmically on the mass of the heavy particle. If this happens, 
effects at low energies can be sizeable. A classical example
is provided by two U(1) gauge fields coupled to a heavy fermion charged with respect to both fields \cite{holdom}. At energies below the
fermion mass $M$, a kinetic mixing of the two gauge fields appears whose strength scales as $\sim \log M$. 

Many BSM scenarios contain a hidden sector of particles coupled to visible degrees of freedom through massive messengers. In these models,
U(1) gauge fields in the hidden sector can have kinetic mixing with a visible U(1), provided there are
messengers charged under the two gauge fields. Moreover, large logarithms at low energies can be resumed using the renormalization group
equation, allowing still for relatively large mixings \cite{bkmr,dkmr} that dominate over other interactions 
mediated by mass-suppressed 
higher-dimensional operators \cite{dobrescu}. Kinetic mixing of the standard model hypercharge with some kind of ``hidden'' U(1)'s emerges
also in many other scenarios. For example, 
with Ramond-Ramond photons in type II string 
compactifications \cite{jl,gkk,gjrr,cim} or in F-theory GUTS \cite{hv}.

Hidden U(1)'s can acquire masses below the eV scale by spontaneous symmetry breaking or the St\"uckelberg mechanism.
They are examples of the class of weakly interacting sub-eV particles (WISPs)  \cite{rev1} that also include axions and axion-like 
particles (ALPs),  as well as minicharged particles. These particles are  very much of interest in cosmology,
both for their possible imprints on the cosmic microwave background (CMB) and as dark matter candidates.

Here we will focus on scenarios with a hidden U(1) mixing with the standard model photon and its effect on the CMB.
The kinetic mixing converts ordinary photons into hidden photons and vice versa \cite{okun},  in a mechanism
very  much resembling  the Mikheev-Smirnov-Wolfenstein (MSW) effect for  the propagation of neutrinos in a medium
\cite{giunti-kim}.
Also in this case, the oscillations between hidden and visible photons are driven by the mass difference of the 
particles.

The present day universe is filled with a thermal background radiation which has a nearly
perfect Planck spectrum, the CMB. Over the last two decades there has been
a tremendous advance in the observations of the CMB, with ground based \cite{act,spt}, balloon \cite{boom} and satellite based
telescopes \cite{cobe,wmap,planck}. These have confirmed the small temperature anisotropies $\frac{\Delta T}{T}\sim {\cal O}(10^{-5})$
and polarization of the CMB. 
The nearly perfect Planck spectrum of the CMB was observed with the COBE/FIRAS instrument \cite{cobe}. 

Spectral distortions encapsulate the deviations from the perfect Planck spectrum \cite{sz2,zeld2, ddz,hs,cs,ks}. 
Mixing between the ordinary photons of the CMB and a WISP of the hidden sector results in a change of the 
number density of photons, thereby causing a spectral distortions of the CMB.
This was  studied for the conversions of photons to hidden photons in \cite{ggg,mrs1}.
For the conversion of photons into ALPs the presence of a magnetic field is required,
which allows to constrain a possible primordial magnetic field \cite{yy,mrs2,hnt,tsm,ed}.

Currently there are several proposals for future missions that have as  one of their goals 
to measure the spectral distortions of the CMB such as PRISM/COrE \cite{prism} 
and PIXIE  \cite{pixie}.
The aim of this work is to use the projected limits on the  CMB spectral distortions from these proposals
to constrain the parameter space of the mixing of photons of the visible and hidden sector.
In section \ref{s2} we provide the key elements to calculate the spectral
distortion of the CMB from photon-hidden photon oscillations.
In section \ref{s3} the constraints on the parameter space of the mixing are provided
and the conclusions are given in section \ref{s4}.

\section{Spectral distortions from mixing of photons of the visible and hidden sectors}
\label{s2}
\setcounter{equation}{0}

The interaction between the standard model photon and a hidden photon of mass $m_{\gamma'}$
leads to additional terms in the low energy action of the form \cite{mrs1,rev1}
\begin{eqnarray}
\Delta{\cal L}_{\gamma\gamma'}=-\frac{1}{4}X_{\mu\nu}X^{\mu\nu}+\frac{\sin\chi_0}{2}X_{\mu\nu}F^{\mu\nu}
+\frac{\cos^2\chi_0}{2}m_{\gamma'}^2 X_{\mu}X^{\mu},
\label{l1}
\end{eqnarray}
where $X_{\mu}$ is the hidden U(1) gauge field with field strength $X_{\mu\nu} =\partial_{\mu}X_{\nu}-\partial_{\nu}X_{\mu}$. 
In addition, $F_{\mu\nu}$  is the 
field strength of the visible U(1) gauge field and $\chi_0$ is the mixing angle.
If the model is embedded in string theory then, e.g., in the case of compactification of the heterotic string
the mixing angle $\chi_0$ has to be in the range between $10^{-17}$ and $10^{-5}$ \cite{dkmr}.
For compactifications in type II string theory $\chi_0$ is in the range $10^{-12}$ to $10^{-3}$ \cite{gjrr}.

The kinetic mixing in the action \eqref{l1} can be undone by a linear field redefinition, resulting in a diagonal kinetic term 
and a nondiagonal mass matrix. In fact, this can be done in such a way that visible matter only couples to one of the new gauge fields, that
we identify with the standard model, interaction eigenstate photon. At the same time, hidden sector matter coupled to $X_{\mu}$ 
gets a coupling to the visible photon suppressed by $\tan\chi_{0}$. This is at the basis of the construction of 
milli-charged dark matter scenarios \cite{fln,ssf}.

The mass matrix can be further diagonalized by a unitary rotation of the two interaction eigenstates that preserves the
diagonal kinetic term. The resulting new gauge fields are the two propagation eigenstates, 
one of which is close to the standard model photon and the other to the hidden photon. This small difference between 
interaction and propagation eigenstates is responsible for the oscillations between hidden and visible
photons.  

However, photons of the CMB do not propagate in vacuum but in the primordial plasma. This leads to birefringence
which can be effectively described by a mass term $m_{\gamma}$
in the dispersion relation. This effective photon mass at a photon energy $\omega$ can be approximated by \cite{mrs1}
\begin{eqnarray}
m_{\gamma}^2\simeq \omega_P^2-2\omega^2({\frak n}-1)_H,
\end{eqnarray}
where $\omega_P=\frac{4\pi\alpha n_e}{m_e}$ is the plasma frequency, $\alpha$ the fine structure constant, and $m_e$ and $n_e$ are the mass and number density of free electrons, respectively. Here $({\frak n} -1)_H=13.6\times 10^{-5}$ is the refractive index of neutral hydrogen under normal
conditions \cite{bw}.
This leads to the following value of the effective photon mass as a function of the redshift $z$ \cite{mrs1}
\begin{eqnarray}
\left(\frac{m_{\gamma}(z)}{\rm eV}\right)^2\simeq 1.4\times 10^{-21}\left[x_e(z)-7.3\times 10^{-3}\left(\frac{\omega(z)}{\rm eV}\right)^2\left(1-x_e(z)\right)
\right]\left(\frac{n_p(z)}{{\rm cm}^{-3}}\right).
\end{eqnarray}
The photon frequency evolves as $\omega(z)=\omega_0(1+z)$ with $\omega_0$ the present day value and the proton number density as
\begin{eqnarray}
n_p(z)=\left(1-\frac{Y_p}{2}\right)\eta\frac{2\zeta(3)}{\pi^2}T_0^3\left(1+z\right)^3,
\end{eqnarray}
for the helium mass fraction $Y_p$, the baryon-to-photon ratio $\eta$  and the CMB temperature today, $T_0$.

\begin{figure}[t]
\centerline{\epsfxsize=3.5in\epsfbox{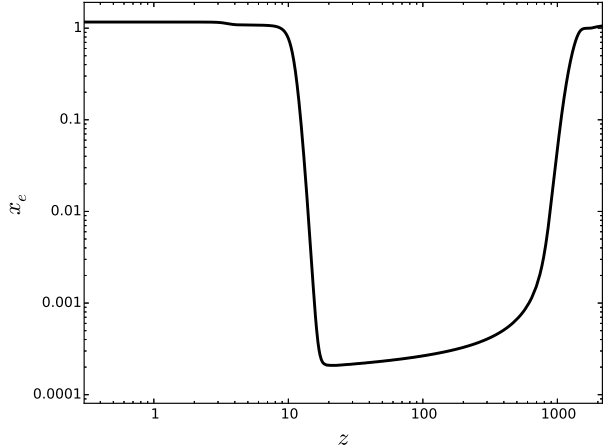}}
\caption{Evolution of the ionization fraction $x_e$ with redshift $z$ calculated with CLASS for the best fit values of Planck 2013 data plus the low $\ell$ polarization data of WMAP 9 \cite{planck,wmap}. }
\label{fig1}
\end{figure}

The oscillation between CMB photons and hidden photons is most efficient when the resonance condition $m_{\gamma}(t_{res})=m_{\gamma'}$ is satisfied at some time $t_{res}$.
The conversion probability $P_{\gamma\rightarrow\gamma'}$ is maximal in the non-adiabatic limit  in which  it
can be approximated by \cite{mrs1}
\begin{eqnarray}
P_{\gamma\rightarrow\gamma'}\simeq\frac{\pi m_{\gamma'}^2\chi_0^2}{\omega}\left|\frac{d\ln m_{\gamma}^2(t)}{dt}\right|^{-1}_{t=t_{res}}.
\end{eqnarray}
This determines the spectral distortion of the CMB since the intensity $I_{\omega}$ will be given by
\begin{eqnarray}
I_{\omega}(m_{\gamma'},\chi_0)=B_{\omega}\left(1-P_{\gamma\rightarrow\gamma'}\right),
\label{e_6}
\end{eqnarray}
where $B_{\omega}=\frac{\omega^3}{2\pi^2}\left[\exp\left(\frac{\omega}{T}\right)-1\right]^{-1}$ is the spectral intensity of  the Planck spectrum.
The conversion probability can be calculated in terms of redshift by expressing the time derivative of the effective photon mass term as a derivative with respect to $z$ and using 
\begin{eqnarray}
\frac{dz}{dt}=-H_0(1+z)\sqrt{\Omega_{\Lambda}+\Omega_m(1+z)^3+\Omega_r(1+z)^4},
\end{eqnarray}
where $\Omega_{\Lambda}$ is the density parameter of dark energy, $\Omega_m$ that of matter and 
the relativistic contribution is determined by $\Omega_{r}=2.471\times 10^{-5}\left(\frac{T_0}{2.725\; {\rm K}}\right)h^{-2}$  \cite{pdg12}.
The effective photon mass depends on the ionization history of the universe which is shown in 
in figure \ref{fig1} for the best fit values of Planck 2013 data plus the low $\ell$ polarization data of WMAP 9
\cite{planck, wmap}. 
The ionization history is calculated with RECFAST \cite{recfast1,recfast2,recfast3,recfast4} as included 
in the  CLASS \cite{class1,class2} code whose purpose actually is the  calculation of  the temperature anisotropies
and polarization of the CMB. Using a code like CLASS allows to take advantage of the already implemented standard reionization
history of the universe. This means that the universe is instantaneously reionized at the reionization redshift, $z_{re}$.
For the best fit values of Planck 2013 as used here, this is given by $z_{re}=11.37$ \cite{planck,wmap}.
The implementation in CLASS uses a tanh fitting formula for the ionization fraction centered at the reionization redshift which is the same
as in CAMB \cite{camb1,camb2}.

\section{Results}
\label{s3}
\setcounter{equation}{0}

\begin{figure}[t]
\centerline{\epsfxsize=3.5in\epsfbox{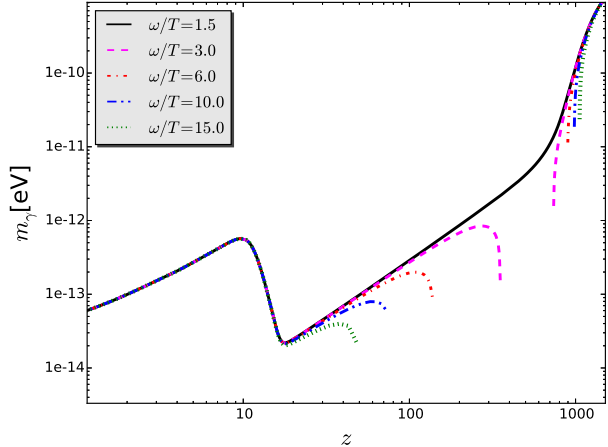}}
\caption{The effective photon mass $m_{\gamma}$ as function of redshift  for the best fit values of Planck 2013 data plus the low $\ell$ polarization data of WMAP 9 \cite{planck,wmap}. }
\label{fig2}
\end{figure}

\begin{figure}[t]
\centerline{\epsfxsize=3.5in\epsfbox{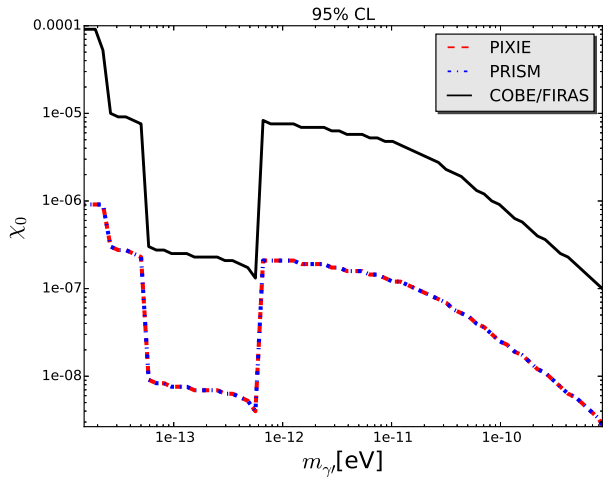}}
\caption{Exclusion plot at 95\% CL derived using the COBE/FIRAS data \cite{cobe} and the expected sensitivities of the planned
experiments PIXIE \cite{pixie} and PRISM \cite{prism}. The cosmological background is determined by the best fit parameters of   
Planck 2013 plus the low $\ell$ polarization data of WMAP 9 \cite{planck,wmap}.}
\label{fig3}
\end{figure}

In the numerical solutions the best fit values of Planck 2013 data plus the low $\ell$ polarization data of WMAP 9
\cite{planck, wmap} are used, $\Omega_{\Lambda}=0.6817$, $\Omega_m=0.3183$, $h=0.6704$, $Y_p=0.24$ and 
$\eta=6.19\times 10^{-10}$. Moreover, for the ionization fraction  $x_e$ we use the numerical solution obtained from 
the CLASS code as shown in figure \ref{fig1}. In figure \ref{fig2} the evolution of the effective photon mass is shown for 
several values of $\frac{\omega}{T}$. The range of values for COBE/FIRAS is $1.2<\frac{\omega_0}{T_0}<11.3$ for the 
present day values. PIXIE is proposed to have  400 frequency channels spanning the frequency range from 30 GHz to 6 THz \cite{pixie}.
Each channel covers a frequency band of $\Delta\nu=15$ GHz. The spectral resolution is $\delta I_{\nu}=5\times 10^{-26}\;{\rm Wm^{-2}Hz^{-1}sr^{-1}}$ \cite{pixie}.
PRISM covers the same frequency range as PIXIE and has simulated global 4-year mission sensitivities, $\delta I_{\nu}=3.6\times 10^{-27}\; {\rm Wm^{-2}Hz^{-1}sr^{-1}}$
in the frequency range $30-180$ GHz, $\delta I_{\nu}=6.5\times 10^{-27}\;{\rm Wm^{-2}Hz^{-1}sr^{-1}}$ in the frequency band 180-600 GHz, $\delta I_{\nu}=7.6\times 10^{-27}\;{\rm Wm^{-2}Hz^{-1}sr^{-1}}$ for frequencies 600-3000 GHz and $\delta I_{\nu}=1.6\times 10^{-26}\;{\rm Wm^{-2}Hz^{-1}sr^{-1}}$ in the frequency band 
3-6 THz \cite{prism}. In each band the resolution is 15 GHz.
Therefore, for PIXIE and PRISM the  range of frequencies implies $0.5<\frac{\omega_0}{T_0}<105.6$.
We follow the same strategy as in \cite{mrs1} and assume that the present day temperature of the CMB is a free parameter which is to be determined by 
minimizing the reduced $\chi^2$ function,
\begin{eqnarray}
\chi^2=\frac{1}{N-1}\sum_i\left[\frac{I^{exp}_i-I_{\omega}(m_{\gamma'},\chi_0)}{\sigma^{exp}_i}\right]^2,
\end{eqnarray}
for each pair of values in parameter space spanned by the mass of the hidden photon $m_{\gamma'}$ and the kinetic mixing parameter $\chi_0$.  In 
the previous expression, $I^{exp}_i$ is the observed or projected spectral intensity and $\sigma^{exp}_i$ the corresponding error.
The model prediction for $I_{\omega}(m_{\gamma'},\chi_0)$ is given by equation (\ref{e_6}).
In figure \ref{fig3} the exclusion region at 95\% confidence limit is shown in the parameter space of
the mass of the hidden photon $m_{\gamma'}$ and the mixing angle $\chi_0$.
As can be seen in this plot, the projected limits of PIXIE and PRISM lead to very similar
constraints on $m_{\gamma'}$ and $\chi_0$. In comparison to the bounds obtained from COBE/FIRAS
observations  future PIXIE-like experiments will improve the bounds on hidden-visible photon interaction
by more than an order of magnitude. This is similar to the improvement found in \cite{ed} for the constraints on the 
combined magnetic field strength and ALP coupling constant  and mass.
Future experiments will be able to constrain the kinetic mixing parameter $\chi_0$ to be less than $10^{-8}$ for the mass of the hidden photon in the range 
$10^{-14}\;{\rm eV}\lesssim m_{\gamma'}\lesssim 10^{-13}\;{\rm eV}$.

\section{Conclusions}
\label{s4}
\setcounter{equation}{0}

There is mounting evidence for the existence of a sector of light particles weakly interacting with the standard model degrees of freedom:
neutrino masses seem to be in the sub-eV range and the energy scale of the cosmological constant is also in this region. 
Different BSM scenarios also contribute to engross this sector with a number of light particles whose interactions with the visible sector are very weak. 
These WISPs are very interesting in cosmology, not only because they could be viable candidates for dark matter, but also because 
of possible detectable effects on the CMB. 

In particular, spectral distortions of the CMB offer the possibility to test fundamental theories of physics. 
Here hidden photons have been considered. They mix with the ordinary standard model photon according to a 
mechanism similar to neutrino oscillations. The effective Lagrangian describing this mixing is
characterized by two parameters, the kinetic mixing parameter $\chi_0$ and the mass of the hidden photon $m_{\gamma'}$.

Extending the analysis of \cite{mrs1}, we have improved the constraints on $\chi_0$ and $m_{\gamma'}$ using the 
expected sensitivities of two proposed experiments, namely, PRISM/COrE \cite{prism} and PIXIE \cite{pixie}. 
For the cosmological background the best fit parameters of Planck 2013 plus the low $\ell$ polarization data of WMAP 9 \cite{planck,wmap} are used.
The characteristic sensitivities of the proposals of PRISM and PIXIE
are very similar. Hence the exclusion  regions in parameter space coincide (cf. figure \ref{fig3}). It is found that for the mass of the hidden photon 
in the range $10^{-14}\;{\rm eV}\lesssim m_{\gamma'}\lesssim 10^{-13}\;{\rm eV}$ the kinetic mixing angle $\chi_0$ has to be less
than $10^{-8}$.  The bounds calculated from the expected sensitivities of PRISM and PIXIE  are shown to be more than an order of magnitude stronger than 
those derived from the COBE/FIRAS data.

\section{Acknowledgements}

KEK would like to thank NORDITA for hospitality during the final stages of this work.
Financial support from Spanish Science Ministry grants FIS2012-30926 and
CSD2007-00042 is gratefully acknowledged.
MAVM  acknowledges financial support from Spanish Science Ministry grants FIS2012-30926, FPA2012-34456, and CSD2007-00042,
as well as from Basque Government Grant IT-559-10.
We acknowledge the use of the Legacy Archive for
Microwave Background Data Analysis (LAMBDA). Support for LAMBDA is
provided by the NASA Office of Space Science.


\bibliography{references}
\end{document}